\documentclass[
superscriptaddress,amsmath,amssymb,aps,twocolumn,
notitlepage,
]{revtex4-2}

\usepackage{appendix}
\usepackage{graphicx}
\usepackage{dcolumn}
\usepackage{bm}
\usepackage{lipsum}
\usepackage[dvipsnames]{xcolor}
\usepackage[T1]{fontenc}
 \fontfamily{cmr}

\usepackage{float}
\usepackage{hyperref}
\usepackage{titlesec}
\usepackage{lipsum}
\usepackage{tikz-cd}
\usepackage{todonotes}
 
\tikzcdset{every label/.append style = {font = \small}}

\titleformat*{\section}{\large\bfseries}
\usepackage[normalem]{ulem}
\hypersetup{
    colorlinks=true,
    linkcolor=blue,
    citecolor=magenta,      
    urlcolor=cyan,
    }

\begin{document}

\preprint{APS/123-QED}

\title{Controlling quantum entanglement with classical non-separable light} 

\author{R. F. Barros}
\email{rafael.fprb@usp.br}

\affiliation{Tampere University, Photonics Laboratory, Physics Unit, Tampere, FI-33720, Finland}
\affiliation{Instituto de Física, Universidade de São Paulo, 05315-970 São Paulo, SP, Brazil}
\author{A. L. S. Santos Junior}%
\affiliation{Instituto de Física, Universidade Federal Fluminense, Niterói, Rio de Janeiro 24210-346, Brazil}
\author{A. Z. Khoury}%
\affiliation{Instituto de Física, Universidade Federal Fluminense, Niterói, Rio de Janeiro 24210-346, Brazil}
\author{R. Fickler}
\affiliation{Tampere University, Photonics Laboratory, Physics Unit, Tampere, FI-33720, Finland}

\date{\today}

\begin{abstract}


Here we investigate the quantum frequency conversion of entangled photons driven by a classically non-separable laser beam. We show that the frequency conversion dynamics is described by a quantum channel that stems from the classical drive field through the channel-state duality - the quantum channel is dual to the classical coherence matrix of the drive field. This implies that the evolution of entanglement in the conversion process is bound by the classical non-separability of the drive field, a result that we confirm experimentally. Furthermore, we show that the conversion dynamics can be understood as a swapping operation between classical non-separability and entanglement, unveiling a physical connection between two fundamentally different concepts.

\newpage
\end{abstract}

\maketitle

\noindent Entanglement is the characteristic trait of quantum mechanics \cite{Schrödinger_1935}. Joint measurements of entangled quantum particles can be perfectly correlated, regardless of how far apart the particles are, while individual measurements lead to random outcomes \cite{EPR,paneru2020entanglement}.
Since its inception and first experimental demonstrations, it not only forced us to reconsider our understanding of the world, but also led a myriad of quantum technological applications driving the second quantum revolution \cite{zhang2024entanglement}.
While no classical system has such a non-local feature, correlations and shared randomness can also be found within the properties of structured optical fields. If the polarisation of a light beam varies across the transverse plane, for example, the beam can seem unpolarized when integrated in space, although its polarisation is perfectly defined at every point \cite{zhan2009cylindrical,forbes2021structured}. This is one example of an optical field that is non-separable in space and polarisation, also called a spin-orbit field.
The reason for that is the association of spin angular momentum with the polarisation state and orbital angular momentum with the transverse field distribution \cite{bliokh2015spin}. Other examples of non-separable fields are spectral vector beams \cite{kopf2021spectral}, for which the polarisation changes across the frequency spectrum, and spatio-temporal vortices~\cite{wan2023optical}, which have transverse structures that change in time. Spatio-spectral vector beams go a step further, coupling space, time, and polarisation at once~\cite{kopf2023correlating}.

All those optical fields have mathematical constructions that are identical to those of entangled quantum states - both are represented by non-factorizable functions in composite vector spaces~\cite{paneru2020entanglement}. This connection between classical non-separability and entanglement allowed for a plethora of quantum-inspired experiments using the degrees of freedom of a single beam of light, such as the violation of Bell's inequality~\cite{Cadu2010}  and local teleportation \cite{Pinheiro2016spin,guzman2016demonstration}, inciting controversial discussions in the literature. Recently, it has been proposed that the quantum-classical distinction is not only the non-locality~\cite{korolkova2024operational}, but more importantly, the number of measurements needed to attest the non-separability.

A promising platform where concepts of quantum and classical optics intersect is that of nonlinear optics. Spontaneous parametric downconversion (SPDC) in non-linear crystals is the standard method for entanglement generation~\cite{couteau2018spontaneous}, but the non-linear process itself follows selection rules involving the spatial~\cite{walborn2010spatial}, temporal~\cite{Huang:13,eckstein2011quantum} and polarisation modes involved~\cite{boyd2008nonlinear}. In a recent experiment, Pinheiro et al.~\cite{Pinheiro2022} demonstrated that in the non-linear interaction of a non-structured (Gaussian) polarised laser beam driven by a spin-orbit field, the information contained in the input polarisation is transferred to the spatial domain as the field changes colour. This result raises the question of how the frequency conversion of a bona fide quantum field is affected by the classical non-separability of the laser beam that drives it.

\begin{figure}
    \centering
    \includegraphics[width=0.8\linewidth]{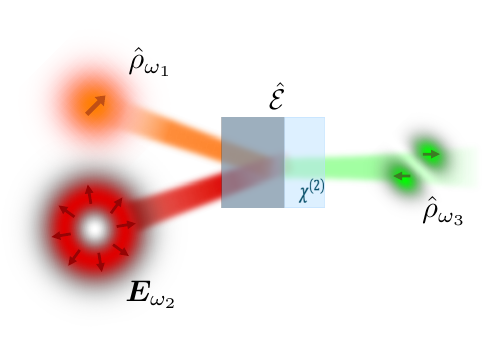}
    \caption{\textbf{Conceptual image of the quantum frequency conversion process}. \textbf{a} An input quantum state $\hat{\rho}_{\omega_1}$ in the polarization of a photon of frequency $\omega_1$ interacts with a non-separable drive field $\boldsymbol{E}_{\omega_2}$ in a non-linear crystal. When upconverted, the input photon changes to the frequency $\omega_3=\omega_1+\omega_2$, and its spatial modes carry the quantum state $\hat{\rho}_{\omega_3}=\hat{\mathcal{E}}(\hat{\rho}_{\omega_1})$.
    }
    \label{Fig_Concept}
\end{figure}

In this article, we report on the theoretical and experimental investigation of the quantum frequency conversion (QFC)~\cite{kumar1990quantum, huang1992observation} driven by classically nonseparable light. 
We show that when the process is used to convert a photon out of an etangled pair it enables to leverage the nonseparability of the classical drive to control the quantum nonseparability of the photon pair.  
In our experiment. we use a pair of photons entangled in polarisation, and drive the upconversion of one of the photons using a strong spin-orbit laser beam. We measure the joint two-colour quantum state with full-state tomography, correlating the spatial mode of the upconverted photon with the polarisation of its original pair. We find that the spin-to-orbit conversion follows an open quantum dynamics, which limits the degree of entanglement of the final quantum state. Interestingly, the quantum channel that describes the dynamics is connected to the coherence matrix of the classical drive field through the state-channel duality~\cite{jiang2013channel}, showing that the spin-orbit nonseparability of the classical drive steers the entanglement dynamics. Lastly, the state-channel duality draws a parallel between the frequency conversion and an entanglement swapping experiment, where the classical drive field emulates a second pair of entangled photons.

\vspace{-1em}

\noindent

\section*{Results and Discussion}

\vspace{-1em}

\noindent\textbf{Fundamentals} 
\vspace{0.1em}

\noindent The QFC process of interest is depicted in Fig. \ref{Fig_Concept}a.
A single photon of central frequency $\omega_1$ is prepared in a polarisation state $\hat{\rho}_{\omega_1}$. The polarization basis modes are chosen as $\boldsymbol{\epsilon}_{1}=\boldsymbol{x}$ and $\boldsymbol{\epsilon}_{2}=\boldsymbol{y}$, which are taken as the ordinary and extraordinary axes of the non-linear crystal, respectively.  The propagation direction is $\boldsymbol{z}$, and the transverse spatial profile in the $xy$ plane is a fundamental Gaussian mode with Rayleigh length $z_R$.

The input photon is focused on a second-order non-linear crystal along with a strong coherent field of central frequency $\omega_2$, which we call the drive field, with the following transverse electric field distribution 
\begin{eqnarray}
\boldsymbol{E}_{\omega_2} (\boldsymbol{r})&=&\sum_{i,j=1}^2 \alpha_{ij}~\boldsymbol{\epsilon}_i S_j(\boldsymbol{r})\,.
\label{EQ_DriveField}
\end{eqnarray} 
The functions $S_1(\boldsymbol{r})$ and $S_2(\boldsymbol{r})$ are a pair of orthonormal spatial modes, and $A=[\alpha_{ij}]$ is an arbitrary complex matrix with unit Frobenius norm ($\sum_{ij}|\alpha_{ij}|^2=1$). In our case, we take $S_1(\boldsymbol{r}) = \textrm{HG}_{10}(\boldsymbol{r})$ and $S_2(\boldsymbol{r})= \textrm{HG}_{01}(\boldsymbol{r})$, where $\textrm{HG}_{mn}(\boldsymbol{r})$ are Hermite-Gaussian spatial modes \cite{Saleh2019} with the same Rayleigh length $z_R$ as the input field. 

The drive field described by Eq.\eqref{EQ_DriveField} is an example of a classically non-separable structure in the polarisation and spatial degrees of freedom (DoF) - its polarisation direction changes across the transverse spatial dimension. Since the spin and orbital angular momenta of light are associated with the polarisation and the transverse spatial structure, respectively, such fields are also called spin-orbit fields~\cite{bliokh2015spin}. In general, correlations between different DoF of a light beam can be assessed through the optical coherence matrix~\cite{kagalwala2015optical}, which in the spin-orbit case is defined as 
\begin{equation}
\rho_D = \boldsymbol{\alpha}\boldsymbol{\alpha}^\dagger,\quad \boldsymbol{\alpha}=(\alpha_{11}~ \alpha_{12}~ \alpha_{21}~ \alpha_{22})^T\,.   
\label{EQ_coherenceMatrix}
\end{equation}
The coherence matrix is the classical-optical analogue of the quantum density operator, and it can be constructed from a tomographically complete set of measurements~\cite{james2001measurement} in the DoF involved. From the coherence matrix we can quantify the degree of non-separability using the concurrence \cite{Hill1997,HorodeckiEntanglement}, which in our case is simply
\begin{equation}
    C(\rho_D) = 2|\textrm{det}(A)|\,.
\end{equation}
Note that the concurrence ranges from $C=0$ for a separable field to $C=1$ for a maximally non-separable field.

Now we address the non-linear interaction between the input photon and the drive field. To find the modes to which the input photons can be converted, we must consider the mode selectivity of the nonlinear interaction. First, energy conservation implies that the central frequency of the upconverted field must be $\omega_3 = \omega_1 + \omega_2$. The polarisation selectivity is affected by two mechanisms, which are the symmetries of the non-linear susceptibility tensor $\chi^{(2)}$, and the phase matching condition, which depend on material dispersion~\cite{boyd2008nonlinear}. The latter, especially, can be designed using ferroelectric domain engineering to allow only selected interactions~\cite{franken1963optical}. Here we consider that the $x+x\rightarrow x$ and $y+y\rightarrow x$ interactions occur with equal efficiency, and that all other interactions are not phase-matched. Although this setting might seem too restrictive, we discuss in the experimental section one way to implement it. Finally, the spatial selection rules imply that only the interactions $\textrm{HG}_{00} + \textrm{HG}_{01} \rightarrow \textrm{HG}_{01}$ and $\textrm{HG}_{00} + \textrm{HG}_{10} \rightarrow \textrm{HG}_{10}$ are possible and that they have the same efficiency~\cite{Alves2018}. In other words, the nonlinear process not only converts the frequency of the photon but at the same time its spatial mode structure, depending on the drive field, while also determining its polarisation.

\begin{figure*}
    \centering
    \includegraphics[width=1\linewidth]{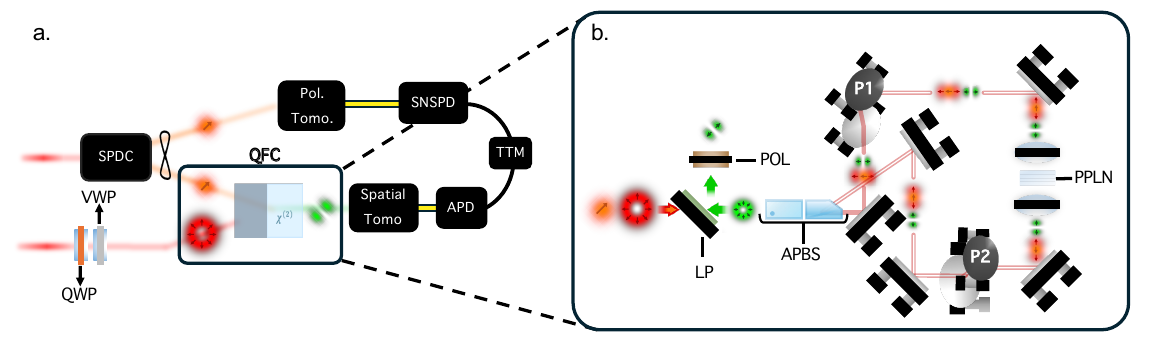}
    \caption{\textbf{Experimental setup.} \textbf{a} Simplified sketch of the experimental setup. A 780~nm pulsed laser is used both as the pump for an SPDC source and as the drive field for Quantum Frequency Conversion (QFC). From the pair of 1560~nm SPDC photons, one photon is measured directly with polarisation projections and a Superconducting Nanowire Detector (SNSPD), while the second photon undergoes Quantum Frequency Conversion (QFC) with the structured drive field. The upconverted 520~nm photon is measured via spatial mode projections and detection with a single photon Avalanche Photodiode Detector (APD). Full state tomography is performed by changing the polarisation and spatial projections and counting coincident detections with a time-tagging module (TTM). \textbf{b} Detailed design of the QFC source, comprising a type-0 non-linear crystal in an achromatic Sagnac loop. \textbf{Legend} HWP, half-wave plate; PBS, polarizing beam-splitter; QWP, quarter-wave plate; VWP, vortex half-wave plate; LP, longpass dichroic mirror; Pol, polarizer; APBS, achromatic polarizing beam splitter; P1 and P2, periscopes; PPLN, periodically-poled lithium niobate crystal.}
    \label{Fig_Setup}
\end{figure*}


\vspace{1em}
\noindent\textbf{Spin-orbit quantum frequency conversion} 
\vspace{0.1em}

\noindent The interaction Hamiltonian that describes the QFC process with all the aforementioned constraints is the following~\cite{kumar1990quantum}
\begin{equation}
    \hat{\mathcal{H}} = i\hbar\kappa \sum_{jk=1}^2 \left(\alpha_{jk}\hat{a}_j^\dagger\hat{b}_k+\textrm{h.c.}\right)\,,
    \label{EQ_Hamiltonian}
\end{equation}
where $\{\hat{a}_j^\dagger,\hat{a}_j\}$ are the creation and annihilation operators of the polarization modes at frequency $\omega_1$, and $\{\hat{b}_k^\dagger,\hat{b}_k\}$ correspond to the spatial modes at frequency $\omega_3$.  The overall strength of the interaction is given by $\kappa \propto \chi^{(2)}_{eff}\sqrt{\bar{n}_D}$, where $\chi^{(2)}$ is an effective non-linear susceptibility and $\bar{n}_D \gg 1$ is the average number of photons of the drive field. Note that Eq.\eqref{EQ_Hamiltonian} is the Hamiltonian of a multiport beam splitter, which in our case connects the polarisation modes of the input frequency to the spatial mode of the upconverted frequency.

An interesting particular case is that of
$\alpha_{ij}=\delta_{ij}/\sqrt{2}$, which represents a maximally non-separable drive field with $C({\rho}_D)=1$. In this case, the Hamiltonian \eqref{EQ_Hamiltonian} becomes $\hat{\mathcal{H}}=(\hat{\mathcal{H}}_{1}+\hat{\mathcal{H}}_{2})/\sqrt{2}$, where $\hat{\mathcal{H}}_{j} = i\hbar\kappa\left(\hat{a}_j^\dagger\hat{b}_j+\textrm{h.c.}\right)\,,$
is the Hamiltonian of a pairwise beam-splitting interaction between the modes $\hat{a}_j$ and $\hat{b}_j$. The result is a QFC dynamics that simply transfers the quantum information, unaltered, from the polarisation at frequency $\omega_1$ to the spatial mode at frequency $\omega_2$, i.e., a spin-orbit frequency conversion. A classical version of this phenomenon has been recently implemented using an intense laser as the input field~\cite{Pinheiro2022}.

In the general case, we calculate the upconverted quantum state as follows
\begin{equation}
    \begin{tikzcd}\hat{\rho}_{\omega_1}\arrow[r, "\hat{\mathcal{H}}", line width = 0.6]&\hat{\rho}(t)\arrow[r, "\hat{P}_{\omega_3}", line width = 0.6]&\hat{\rho}_{\omega_3}\,.
    \end{tikzcd}
    \label{EQ_Evolution}
\end{equation}
First, the input state $\hat{\rho}_{\omega_1}$ evolves for an interaction time $t$ into a two-color quantum state $\hat{\rho}(t)$ defined by the Hamiltonian \eqref{EQ_Hamiltonian}. Then, we project $\hat{\rho}(t)$ onto the subspace of modes with frequency $\omega_3$ using the projection operator $\hat{P}_{\omega_3}=\sum_{j=1}^2\hat{b}^\dagger_j|0\rangle\langle 0 |\hat{b}_j$, obtaining 
\begin{equation}
    \hat{\rho}_{\omega_3} = \frac{\hat{P}_{\omega_3}\hat{\rho}(t)\hat{P}_{\omega_3}}{\textrm{Tr}\{\hat{P}_{\omega_3}\hat{\rho}(t)\hat{P}_{\omega_3}\}}\,,
    \label{EQ_Projection}
\end{equation}
which is the quantum state describing spatial modes at frequency $\omega_3$.
As we detail in Appendix~\ref{Theoretical Modelling of Spin-orbit quantum frequency conversion}, the resulting dynamics is catured by a superoperator $\hat{\mathcal{E}}$ defined as 
\begin{equation}
    {\rho}_{\omega_3}=\hat{\mathcal{E}}({\rho}_{\omega_1}) =K^\dagger{\rho}_{\omega_1}K\,,
    \label{EQ_OpenDynamics}
\end{equation}
where ${\rho}_{\omega_1}$ and ${\rho}_{\omega_3}$ are the matrix representations of $\hat{\rho}_{\omega_1}$ and $\hat{\rho}_{\omega_3}$, respectively, and 
\begin{equation}
    K^\dagger = A \left(\sqrt{AA^\dagger}\right)^{-1}\!\!\!\!\sin(\kappa t\sqrt{AA^\dagger})\,.
    \label{EQ_KMatrix}
\end{equation}
Since $K$ is generally not a unitary, we see that $\hat{\mathcal{E}}$ generally describes an open quantum dynamics that is controlled by the drive field through the matrix $A$.

The quantum channel $\hat{\mathcal{E}}$ can be interpreted in terms of the Choi–Jamiołkowski isomorphism, also known as channel-state duality \cite{jiang2013channel}. The single-qubit channel $\hat{\mathcal{E}}$ can be represented by a quantum state
\begin{equation}
    \hat{\rho}_{\mathcal{E}} = (\hat{\mathcal{I}}\otimes\hat{\mathcal{E}})(|\Phi^+\rangle\langle\Phi^+|)\,,
    \label{EQ_ChoiMatrix}
\end{equation}
also called the Choi state. It corresponds to the density operator of a two-qubit system, initially prepared in the maximally entangled Bell state $|\Phi^+\rangle$, after the action of $\hat{\mathcal{E}}$ on one of the qubits. Although the Choi state of $\hat{\mathcal{E}}$ has an intricate dependence on the drive field (See Appendix~\ref{Theoretical Modelling of Spin-orbit quantum frequency conversion}), a surprisingly intuitive result emerges in the limit of low conversion efficiency ($\kappa t\ll 1$). If $\rho_{\mathcal{E}}$ is the matrix representation of 
$\hat{\rho}_{\mathcal{E}} $, we find
\begin{equation}
\begin{tikzcd}
\rho_{\mathcal{E}} \arrow[r, "\kappa t\ll 1", line width = 0.6]&\rho_D\,.
\end{tikzcd}
    \label{EQ_ChoiConnection}
\end{equation}
This is the central theoretical result of this work - in the low efficiency regime, the Choi matrix describing the QFC dynamics \textit{is} the coherence matrix of the classical field used to drive it.

\begin{figure*}
    \centering\includegraphics[width=1\linewidth]{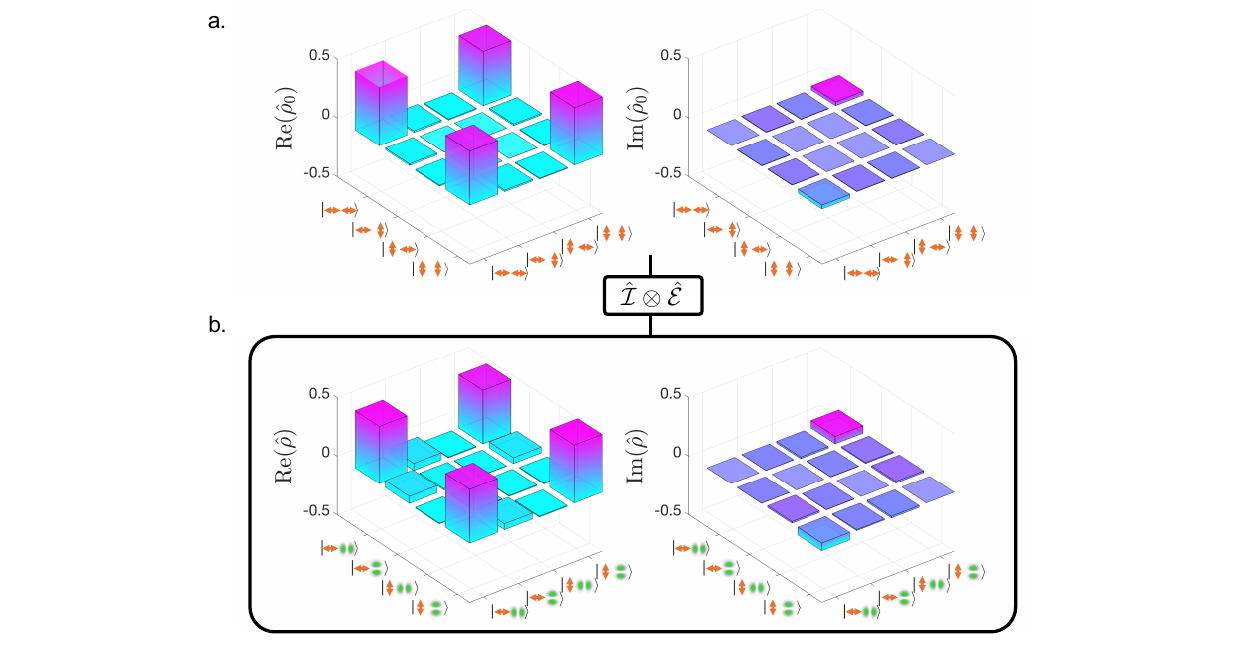}
    \caption{\textbf{Spin-orbit quantum state transfer. } \textbf{a} Quantum state $\hat
    {\rho}_0$ of the SPDC source, obtained via maximum likelihood estimation from a tomographically complete set of polarisation measurements. \textbf{b} Two-color quantum state $\hat
    {\rho}$ obtained after QFC with a maximally non-separable drive field prepared by setting $\theta=0$. The obtained state $\hat
    {\rho}$ has a fidelity of $(97.2\pm0.8)\%$ to the initial state $\hat
    {\rho}_0$, and its concurrence is $\mathcal{C}(\hat
    {\rho})=0.93\pm0.02$.
    }
    \label{Fig_Entanglement}
\end{figure*}

The connection of Eq.\eqref{EQ_ChoiConnection} raises the question of the extent to which the classical correlations contained in the drive field $\rho_D$ can be inherited by a genuine quantum system. Importantly, what happens to the entanglement of a pair of photons in a state $\hat{\rho}_{0}$ when one photon undergoes the QFC dynamics of Eq.\eqref{EQ_ChoiMatrix}? From the theory perspective, this question is answered by a result of Konrad et al.\cite{konrad2008evolution}, which determines that the final two-colour state $\hat{\rho}$ is such that 
\begin{equation}
\begin{tikzcd}
    C(\hat{\rho})\leq C(\hat{\rho}_{\mathcal{E}})C(\hat{\rho}_0) \arrow[r, "\kappa t \ll 1 ", line width = 0.6] &     C(\rho_D) C(\hat{\rho}_0)\,.
\end{tikzcd}
\end{equation}
This result implies that the QFC of an entangled state $\hat{\rho}_0$ leads to a two-colour state $\hat{\rho}$ whose entanglement is bounded from above by the non-separability of the drive field. In the next section, we detail the experimental investigation of this result.

\vspace{1em}
\noindent\textbf{Experiment}

\noindent
The experimental apparatus we use is depicted in Fig.\ref{Fig_Setup}a. We generate a polarisation-entangled pair of photons centred at 1560~nm via spontaneous parametric down-conversion (SPDC). We use a non-colinear Sagnac loop similar to that of Ref.~\cite{kim2019pulsed}, employing a 10 mm-long type-I Lithium Niobate crystal (PPLN, HC Photonics). The pump field derives from the second-harmonic of a mode-locked laser centered at 1560~nm (NKT Origami, 220~fs pulse duration, 80\,MHz, 100\,mW average power) in a 1 mm-long type-0 PPLN crystal (HC Photonics), resulting in a 780nm laser beam with 37 mW average power and negligible temporal walk-off. Only 2mW are used for the pump field, and the remaining 35mW are used to prepare the drive field for QFC. The polarisation state $\rho_0$ of the photon pair is characterised by performing full quantum state tomography~\cite{james2001measurement,altepeter2005photonic}, based on controllable polarisation projections and detection with superconducting nanowire single-photon detectors (SNSPDs, Single Quantum).

Because we use a pulsed drive field, care must be taken to ensure that the photon pairs are not entangled in frequency. This is the reason for using the type-I interaction $1560\,\textrm{nm} (o)+1560\,\textrm{nm}(o)\leftrightarrow 780\,\textrm{nm}(e)$ in PPLN, as the group velocity matching between the pump and the SPDC photons at these wavelengths~\cite{yu2002broadband,horoshko2024few} enhances the phase-matching bandwidth and allows for spectrally pure photons without tight spectral filtering \cite{yu2002broadband,horoshko2024few}. In our case, assuming that both the pump and drive fields have Gaussian pulse profiles with $220$~fs duration, we estimate that the SPDC photons after a 12~nm bandpass filter have $85.9\%$ spectral purity and $92.1\%$ overlap with the pump/drive temporal waveforms (See Appendix~\ref{Requirements for pulsed upconversion}). The high temporal overlap between the drive field and the input photon is essential for optimal QFC efficiency in the pulsed regime.

To imprint a spin-orbit structure on the drive field, we use a quarter-wave plate~(QWP) followed by a vortex half-wave plate~(VWP, Thorlabs WPV10L-1550). The QWP changes the ellipticity of the initial polarisation state, which is initially set to $\boldsymbol{x}$, and the VWP transforms each field component as 
\begin{equation}
\begin{aligned}
    \textrm{HG}_{00}\boldsymbol{x} &\rightarrow \textrm{HG}_{10}\boldsymbol{x} + \textrm{HG}_{01}\boldsymbol{y}\, \\
    \textrm{HG}_{00}\boldsymbol{y} &\rightarrow \textrm{HG}_{01}\boldsymbol{x} - \textrm{HG}_{10}\boldsymbol{y}\,.
    \end{aligned}
    \label{EQ_VortexPlate}
\end{equation}
Therefore, a linearly polarised input becomes a maximally non-separable field, while a circularly polarised input becomes an optical vortex \cite{bliokh2015spin} with uniform circular polarisation. Hence, by tuning the angle $\theta$ between the fast axis of the QWP and the $\boldsymbol{x}$ direction, we generate a drive field whose concurrence is $C(\theta)=|\cos(2\theta)|$, which is the degree of circular polarization before the VWP.

The second part of the setup is the QFC source, detailed in Fig.~\ref{Fig_Setup}b. It consists of a 1mm long type-0 PPLN crystal, phase-mached for $780\,\textrm{nm}(e)+1560\,\textrm{nm}(e)\leftrightarrow520\,\textrm{nm}(e)$ interaction, inside an achromatic sagnac loop~\cite{hentschel2009three}. The frequency conversion goes as follows. The drive field is set to co-propagate towards the Sagnac loop with one of the SPDC photons, which serves as the input photon for QFC, while the second SPDC photon is measured directly with a chosen polarisation projection. 
In the clockwise loop direction, the fields undergo a $90^\circ$ polarisation rotation by a periscope before interacting in the non-linear crystal. In the counterclockwise direction, the fields first interact in the nonlinear crystal, and the upconverted field has its polarisation rotated before exiting the loop. Finally, we project the polarisation of the upconverted $520$~nm photon with a polariser oriented along $(\boldsymbol{x}+\boldsymbol{y})/\sqrt{2}$, erasing the distinguishability of the two loop directions at the cost of limiting the conversion efficiency to $50\%$. This source has the intended mode selectivity and leads to a QFC dynamics described by Eq.\eqref{EQ_OpenDynamics}.

Finally, the spatial mode tomography of the upconverted photon is done via phase-flattening with a phase-only spatial-light modulator (SLM, Holoeye Pluto 2.1) followed by the coupling to a single-mode optical fibre \cite{bouchard2018measuring}. The photons are detected with an avalanche photodiode single-photon detector (APD, Laser Components Count-T) and counted in coincidence with the heralding photon using a time-tagger (TTM, ID Quantique ID900). The coincidence measurements allow for the full quantum state tomography of the two-qubit state $\hat{\rho}$ composed by the polarisation of the heralding 1560~nm photon and the spatial mode of the upconverted 520~nm photon.

\vspace{1em}
\noindent\textbf{Results}

\noindent
We start by generating the quantum state $\hat{\rho}_0$ of the SPDC source shown in Fig.~\ref{Fig_Entanglement}a, with a fidelity of $(95.8\pm0.08)\%$ to the $|\Phi^+\rangle$ Bell state, and a concurrence of $C(\hat{\rho}_0)=0.919\pm0.002$. 
The average pair rate is 2.6~kHz, and each polarisation projection is measured with a 1~min integration time. The standard deviations for the state metrics are obtained via Monte-Carlo simulations assuming Poisson statistics of the photon counts.

Initially, we set the QWP angle to $\theta=0$ to observe the spin-to-orbit quantum frequency conversion. As in this case $\hat{\mathcal{E}}(\hat{\rho}_{\omega_1})=\hat{\rho}_{\omega_1}$, we expect to measure a hybrid two-color entangled state $\hat{\rho}$ with high fidelity to $\hat{\rho}_0$. Experimentally, we obtain a state $\hat{\rho}$ with $(97.2\pm0.8)\%$ fidelity to $\hat{\rho}_0$, and concurrence $C(\hat{\rho})=0.93\pm0.02$, matching the concurrence of the input state within the standard deviation. The density matrix of the upconverted quantum state is shown in Fig.~\ref{Fig_Entanglement}b. To further confirm the hybrid entanglement between the heralding photon and the upconverted photon, we test the direct violation of the Bell-CHSH inequality~\cite{CHSH} (See Appendix~\ref{Violation of Bell's inequality}). We do so by fixing the polarization projections of the heralding photons and rotating the hologram that projects the spatial mode of the upconverted photon, obtaining a maximum Bell-CHSH value of $|\mathcal{B}(\hat{\rho})|_{\textrm{max}}=2.76\pm0.11$ and violating the CHSH inequality by more than $6$ standard deviations.

\begin{figure}
    \centering
\includegraphics[width=1\linewidth]{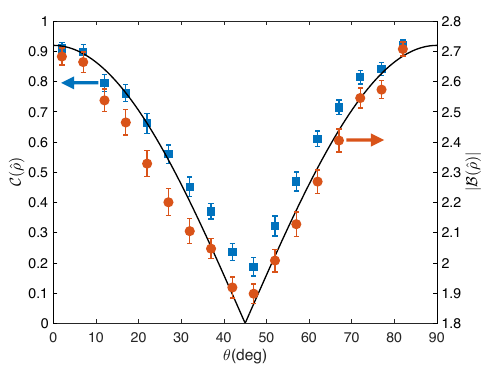}
    \caption{\textbf{Entanglement measurements.} Concurrence and maximum Bell-CHSH value of the two-color quantum state $\hat{\rho}$ for a varying drive field. In the figure, square blue markers refer to the concurrence values on the left axis, and the round orange markers refer to the maximum Bell-CHSH values on the right axis. The continuous black line shows the curve $C(\hat{\rho}_0)|\cos(2\theta)|$.}
    \label{Fig_Concurrence}
\end{figure}

To investigate the connection between the spin-orbit non-separability of the drive field and the entanglement of the upconverted state, we repeat the experiment for different values of the angle $\theta$. For each angle, we perform full state tomography and compute the concurrence and the maximum Bell-CHSH value~\cite{horodecki1995violating} of the state $\hat{\rho}$, obtaining the result shown in Fig.~\ref{Fig_Concurrence}. Our results confirm that $C(\hat{\rho})$ follows the curve $C(\hat{\rho}_0)|\cos(2\theta|$, in good agreement with our theoretical model. Deviations from the model are stronger near $\theta=45^\circ$, where we obtain a minimum concurrence of $C(\hat{\rho}) = 0.19\pm0.03$ and where a vanishing concurrence is expected. We associate this effect with minor experimental imperfections in the preparation of the drive field with the VWP, which lead to a polarisation distribution that is not completely uniform, i.e., still slightly nonseparable, when the input polarisation is circular. However, we note that for the two points of lowest concurrence, the maximum Bell-CHSH does not suffice to violate the Bell-CHSH inequality, showing that although some entanglement persists, the non-locality of the state is lost. These results confirm that the non-separability of the drive field controls the degree of quantum correlations that can be transferred in quantum frequency conversion.

\vspace{1em}
\noindent\textbf{Connection to entanglement swapping}
\noindent

\begin{figure*}
    \centering
    \includegraphics[width=1\linewidth]{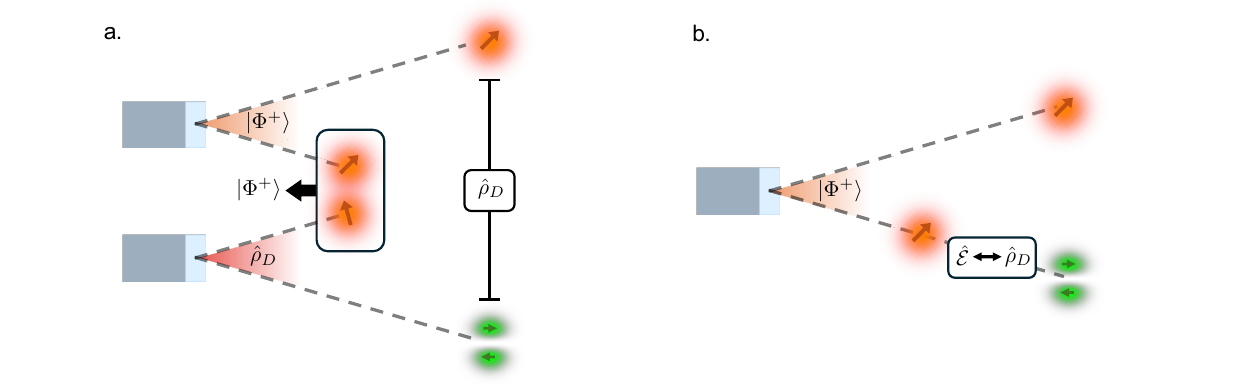}
    \caption{\textbf{Correspondence to entanglement swapping.} \textbf{a} Entanglement swapping experiment. One SPDC source produces a polarisation-entangled pair of photons in the state $|\Phi^+\rangle$, while a separate source produces a second pair of photons in the hybrid entangled state $\hat{\rho}_D$. By projecting one photon from each source on a $|\Phi^+\rangle$ state, the remaining photons are projected onto the entangled state $\hat{\rho}_D$. \textbf{b} An equivalent picture in terms of a single-qubit channel $\mathcal{E}$, dual to the state $\hat{\rho}_D$, acting on one of the photons in a polarization-entangled state $|\Phi^+\rangle$.}
    \label{Fig_Swapping}
\end{figure*}

An interesting analogy can be contained by exploring the connection between the one-sided action of a quantum channel and an entanglement swapping experiment~\cite{konrad2008evolution,pan1998experimental}. Explicitly, consider that an SPDC source produces a pair of photons of frequency $\omega_1$, A and B, sharing a polarisation-entangled state $|\Phi^+\rangle$. In addition, consider a separate SPDC source that produces photons of frequencies $\omega_1$ and $\omega_3$, which we call C and D, respectively, such that the polarization of photon C and the spatial mode of photon D share the quantum state $\hat{\rho}_{D}$. An entanglement swapping experiment using the two SPDC sources is done by projecting the photons B and C in one of the four Bell states, leaving the photons A and D in an entangled state without them ever interacting directly. Particularly, if the photons B and C are projected onto the Bell state $|\Phi^+\rangle$, the photons A and D are projected onto the state $\hat{\rho}_{D}$, i.e., the same quantum state as the second SPDC source.

The dynamics detailed above also describes the action of a one-sided quantum channel~\cite{konrad2008evolution}, as illustrated in Fig.~\ref{Fig_Swapping}. In our particular case, the photon B undergoes a QFC interaction described by the quantum channel $\hat{\mathcal{E}}$, leading to a final two-colour quantum state given by  $\hat{\rho}_{\hat{\mathcal{E}}}=\hat{\rho}_{D}$, the coherence matrix of the drive field (see Eq.\eqref{EQ_ChoiConnection}). Therefore, we see that the QFC process is analogous to an entanglement swapping experiment, where the role of the second SPDC source is played by the spin-orbit coherence matrix of the drive field. In a sense, the polarisation-selective QFC interaction plays the role of a Bell measurement, swapping the spin-orbit non-separability of the drive field with the polarisation entanglement of the SPDC photons. We anticipate that different polarisation settings for the non-linear interaction correspond to different Bell measurements, but a deeper investigation on this subject will be left for future work.


In this article, we investigated the quantum frequency conversion process driven by a spin-orbit field, i.e., a light field whose polarisation is non-uniformly distributed in space. We developed a theoretical framework for spin-orbit QFC, demonstrating that in the low-efficiency regime, the conversion dynamics is governed by a quantum channel that is dual to the optical coherence matrix of the drive field. For an input photon entangled with a second photon, we showed that the entanglement of the final two-colour state is bounded from above by the classical spin-orbit non-separability of the drive field. We experimentally confirmed this result by converting a polarisation-entangled photon pair from 1560~nm to 520~nm using a 780~nm drive field. Furthermore, we showed that the dynamics of QFC driven by non-separable light is mathematically analogous to an entanglement swapping experiment, where the coherence matrix of the drive field mimics the quantum state of a second pair of photons.

Importantly, our results are not limited to the case of spin-to-orbit QFC. A drive field that is non-separable in any two degrees of freedom will induce similar quantum dynamics, provided that the non-linear interaction is selective in one of them. For example, three-wave mixing is known to be selective to the temporal modes involved~\cite{Huang:13}, and spatial mode selectivity can be achieved with appropriate mode projections~\cite{walborn2010spatial}. In this context, hybrid QFC is a promising platform for quantum information processing, especially in DoF that are hard to measure directly. Moreover, the open quantum dynamics of hybrid QFC unveils the limitations in the upconversion of entangled states. This can be particularly important in the QFC of multiple qubits, in which case the entanglement might feature more intricate evolution, such as sudden death and rebirth of entanglement~\cite{almeida2007environment,yu2009sudden}  and non-Markovian dynamics~\cite{breuer2016colloquium}. In summary, our work contributes to a better understanding of quantum frequency conversion and the interplay between quantum entanglement and classical non-separability, stimulating new developments in quantum optics and quantum information science.






\section*{Acknowledgments}

\vspace{-1em}

We thank Kaled Dechoum for the valuable discussions on the quantum frequency conversion dynamics, and Jaime Moreno for the assistance with the experiment. We also thank Meron Tesfa for the characterisation of the vortex plate. R. B. acknowledges the support of the Academy of Finland through the postdoctoral researcher funding (decision 349120). A.K. and A.J. acknowledge the support of Fundação de Amparo à Pesquisa do Estado de São Paulo (FAPESP, grants 2021/06823-5 and 2022/15036-0). R.F. acknowledges the support of the Research Council of Finland through the Academy Research Fellowship (decision 332399). R.B. and R.F. acknowledge the support of the Research Council of Finland through the Photonics Research and Innovation Flagship (PREIN - decision 346511).

\section*{Author contributions}

\vspace{-1em}

\noindent
The experiment was conceived by R.B., R.F. and A.K., driven by A.K.'s idea of studying the quantum counterpart of Ref.~\cite{Pinheiro2022}. The theoretical study was conducted by R.B. and A.J., the experimental setup and measurements were performed by R.B. and A.J., and the data analysis was performed by R.B. The manuscript was written by R.B. and A.J., and edited by all authors. R.F. and A.K. supervised and assisted at every stage of the study.

\section*{Competing interests}

\vspace{-1em}

\noindent
The authors declare no competing interests

\onecolumngrid

\pagebreak
\appendixpage

\section{Theoretical Modelling of Spin-orbit quantum frequency conversion}
\label{Theoretical Modelling of Spin-orbit quantum frequency conversion}

\subsection{Qubit Dynamics Under a Spin-Orbit QFC Hamiltonian}
\label{Qubit Dynamics Under a Spin-Orbit QFC Hamiltonian}

Let us start by writing the drive field in the form
\begin{equation}
\mathbf{E}_{\omega_2}(\mathbf{r})
  = \sum_{i,j=1}^{2} \alpha_{ij}\,
    \boldsymbol{\epsilon}_{i}\, S_{j}(\mathbf{r})\,,
\label{eq:DriveField}
\end{equation}
 where \(S_{1}(\mathbf{r}) = \mathrm{HG}_{10}(\mathbf{r})\) and
\(S_{2}(\mathbf{r}) = \mathrm{HG}_{01}(\mathbf{r})\) constitute an
orthonormal pair of Hermite–Gaussian (HG) spatial modes.
The complex coefficients \(\alpha_{ij}\) form the matrix
\(A = [\alpha_{ij}]\) and satisfy the normalisation
condition \(\sum_{i,j} |\alpha_{ij}|^{2} = 1\). A possible representation for the drive field is given by the coherence matrix~\cite{kagalwala2015optical},
\begin{equation}
\label{eq:CoherenceDrive}
\rho_D = \lvert\alpha)(\alpha\rvert\,,\quad \lvert\alpha) =(\alpha_{11}~ \alpha_{12}~ \alpha_{21}~ \alpha_{22})^T\,,
\end{equation}
in terms of which the correlations between degrees of freedom (DoF) can be conveniently described.  For instance, the degree of non-separability between the polarisation and spatial DoF is quantified by the concurrence~\cite{HorodeckiEntanglement}
\begin{equation}
\label{eq:ConcurrenceDrive}
C(\rho_{D}) = 2\,\bigl|\det (A)\bigr|
            = 2\,\bigl|\alpha_{11}\alpha_{22} - \alpha_{12}\alpha_{21}\bigr| 
            \equiv C_{D},
\end{equation}
which ranges from \(C_{D}=0\) for a separable field to \(C_{D}=1\)
for a maximally non-separable field. We adopt the convention to represent column (row) vectors with curly kets (bras). Square matrices will be represented by the corresponding operator symbol without the hat.

In quantum frequency conversion (QFC), assuming the particular polarisation selectivity that we discuss in the main text, the spin-orbit drive field of Eq.~\eqref{eq:DriveField} leads to the following interaction Hamiltonian 
\begin{equation}
    \hat{\mathcal{H}}
      = i\hbar\kappa
        \sum_{j,k=1}^{2}
        \bigl(
          \alpha_{jk}\,\hat{a}_{j}^{\dagger}\hat{b}_{k}
          + \mathrm{h.c.}
        \bigr),
    \label{EQ_Hamiltonian}
\end{equation}
where \(\kappa\) is an effective nonlinear coupling constant,
\(\{\hat a_{j}^{\dagger},\hat a_{j}\}\) describe the input polarisation
modes, and \(\{\hat b_{k}^{\dagger},\hat b_{k}\}\) the up-converted
spatial modes. The coefficients \(\alpha_{jk}\) thus imprint the spin–orbit structure of the
drive field~\eqref{eq:DriveField} onto the conversion process. Defining vectors $\boldsymbol{\hat{a}}^\dagger = \begin{pmatrix} \hat{a}_{0}^\dagger & \hat{a}_{1}^\dagger \end{pmatrix} \,\, \text{and}\,\, \boldsymbol{\hat{b}}^\dagger = \begin{pmatrix} \hat{b}_{0}^\dagger & \hat{b}_{1}^\dagger \end{pmatrix}$, we find from Eq.~\eqref{EQ_Hamiltonian} the following equations of motion
\begin{equation}
\begin{cases}
\frac{d}{dt}\boldsymbol{\hat{a}}^\dagger = \kappa\,\boldsymbol{\hat{b}}^\dagger A^\dagger\,, \\[6pt]
\frac{d}{dt}\boldsymbol{\hat{b}}^\dagger = -\kappa\,\boldsymbol{\hat{a}}^\dagger A\,, \\[6pt]
\end{cases}
\end{equation}
which have the following general solution
\begin{equation}
\label{abevolution}
\begin{aligned}
\boldsymbol{\hat{a}}^\dagger(t) &= \boldsymbol{\hat a}^\dagger(0)\;\cos\Bigl(\kappa t\,\sqrt{A\,A^\dagger}\Bigr)
+\;\boldsymbol{\hat b}^\dagger(0)\; A\,(A\,A^\dagger)^{-1/2}\sin\Bigl(\kappa t\,{\sqrt{A\,A^\dagger}}\Bigr)\,,\\[2ex]
\boldsymbol{\hat b}^\dagger(t) &= \boldsymbol{\hat b}^\dagger(0)\;\cos\Bigl(\kappa t\,\sqrt{A^\dagger\,A}\Bigr)\,
-\; \boldsymbol{\hat a}^\dagger(0)\;A^\dagger\,(A^\dagger\,A)^{-1/2}\sin\Bigl(\kappa t\,\sqrt{A^\dagger\,A}\Bigr)\,,
\end{aligned}
\end{equation}
where $t$ is the duration of the nonlinear interaction.

We now turn to the time evolution of a polarisation qubit of frequency $\omega_1$ as it evolves under the Hamiltonian of Eq.~\eqref{EQ_Hamiltonian}. We write a general initial state in the following compact notation
\begin{equation}
    \hat{\rho}_{\omega_1} = \sum_{m,n=0}^1 \left(\rho_{\omega_1}\right)_{mn}\,\hat{a}_m^\dagger(0)\,|0\rangle\langle 0|\,\hat{a}_n(0)\equiv \hat{\boldsymbol{a}}^{\dagger}(0)|0\rangle\,\rho_{\omega_1}\,\langle 0|\hat{\boldsymbol{a}}(0)\,,
  \end{equation}
where the density matrix and the vectors of operators are multiplied by standard matrix multiplication.
Using Eqs.~\eqref{abevolution}, we obtain the time-evolved density operator
\begin{equation}
\label{rhot}
\begin{aligned}
\hat{\rho}(t) &=\hat{\boldsymbol{a}}^{\dagger}(t)|0\rangle\,\rho_{\omega_1}\,\langle 0|\hat{\boldsymbol{a}}(t)\\
& = \Bigl[\,
      \boldsymbol{\hat a}^{\dagger}(0)\;\cos\bigl(\kappa t\,\sqrt{A\,A^\dagger}\bigr)\,
      +
      \boldsymbol{\hat b}^{\dagger}(0)\; A\,(A\,A^\dagger )^{-1/2}\,
      \sin\bigl(\kappa t\,\sqrt{A\,A^\dagger}\bigr)\, 
  \Bigr]\,
  |0\rangle\,\rho_{\omega_1}\,\langle 0|
\\[4pt]
&\times
  \Bigl[\,
      \cos\bigl(\kappa t\,\sqrt{A A^\dagger}\bigr)\;\boldsymbol{\hat a}(0)+\,\sin\bigl(\kappa t\,\sqrt{A\,A^\dagger }\bigr)\,
      (A\,A^\dagger)^{-1/2}\,A^\dagger\;\boldsymbol{\hat b}(0)
  \Bigr]\,,
\end{aligned}
\end{equation}
which fully captures the dynamics of quantum frequency conversion driven by a spin-orbit drive field. In general, Eq.~\eqref{rhot} describes a two-colour quantum state, with quantum information encoded in the polarisation modes of frequency $\omega_1$ and the spatial modes of frequency $\omega_3$. However, if only the upconverted modes of frequency $\omega_3$ are measured, as is the case in our experiment, the quantum state that describes the measurements is given by
\begin{equation}
    \hat{\rho}_{\omega_3} = \hat{P}_{\omega_3}\hat{\rho}(t)\hat{P}_{\omega_3}\,,
    \label{EQ_Projection}
\end{equation}
up to a normalisation factor, where $\hat{P}_{\omega_3} = \hat{\boldsymbol{b}}^{\dagger}|{0}\rangle\langle{0}|\,\hat{\boldsymbol{b}}$ is a projection operator. From Eq.~\eqref{rhot} we readily obtain that
\begin{equation}
\label{quantumchannel}
  \hat\rho_{\omega_3} = \hat{\boldsymbol{b}}^{\dagger}(0)|0\rangle\,K^\dagger\rho_{\omega_1}K\,\langle 0|\hat{\boldsymbol{b}}(0)\,,
\end{equation}
where
\begin{equation}
\label{Kmatrix}
    K^\dagger = A\,(A\,A^\dagger )^{-1/2}\,
      \sin\bigl(\kappa t\,\sqrt{A\,A^\dagger}\bigr).
\end{equation}
Therefore, the matrix representation of the upconverted quantum state, now in the basis of spatial modes of frequency $\omega_3$, is
\begin{equation}
\rho_{\omega_3} =  {\mathcal{E}}(\rho_{\omega_1})\equiv K^\dagger\,\rho_{\omega_1}\,K\,,
\label{EQ_Channel_Raw}
\end{equation}
which is positive semi-definite by construction. The superoperator $\mathcal{E}$ defines a completely positive map that is generally not unitary and not trace-preserving, as $\textrm{Tr}\!\bigl[K^\dagger K] =\textrm{Tr}[\,\sin^2(\kappa\sqrt{A A^\dagger} \,t)\,]\le 1$. Therefore, it represents an open quantum dynamics, as we discuss in the following.

\subsection{Entanglement Evolution and channel-state duality}
\label{Entanglement Evolution and channel-state duality}

A convenient way to represent a quantum channel is through the so-called channel–state duality, a fundamental result of quantum information theory that establishes a one-to-one correspondence between completely positive maps and density operators. This correspondence is formalized by the Choi–Jamiołkowski isomorphism \cite{jamiolkowski1972linear,choi1975completely}, which associates each quantum channel \(\hat{\mathcal{E}}\) with a density operator \(\hat{\rho}_{\mathcal{E}}\) called the Choi state. For a single-qubit channel, this state is obtained by applying \(\hat{\mathcal{E}}\) to one subsystem of a maximally entangled two‐qubit Bell state \(\lvert \Phi^+ \rangle\), while leaving the other subsystem untouched. Formally, the Choi state is defined as
\begin{equation}
\label{choimatrix}
      \hat{\rho}_{\mathcal{E}}
  = (\mathcal{I}\otimes\hat{\mathcal{E}})\lvert\Phi^+\rangle\langle\Phi^+\rvert.
\end{equation}
The channel-state duality is particularly useful in quantifying the effect of a quantum channel on the evolution of entanglement in bipartite systems, as discussed in Ref.~\cite{konrad2008evolution}. 
Consider the quantum state $\hat{\rho}= (\mathcal{I} \otimes \hat{\mathcal{E}})\,\hat{\rho}_{0}$, resulting from the action of the quantum channel $\hat{\mathcal{E}}$ on one of the qubits initially in the state $\hat{\rho}_{0}$. In this scenario, Konrad et al.~\cite{konrad2008evolution} showed that the concurrences~\cite{Hill1997,HorodeckiEntanglement} of the initial and final states are related by the following inequality
\begin{equation}
\label{Konrad}
  C(\hat{\rho}) \le C(\hat{\rho}_{\mathcal{E}})\,C(\hat{\rho}_{0}),
\end{equation}
where $C(\hat{\rho}_{\mathcal{E}})$ is the concurrence of the Choi state \( \hat{\rho}_{\mathcal{E}} =(\mathcal{I}\!\otimes\!\hat{\mathcal{E}})\lvert\Phi^+\rangle\langle\Phi^+\rvert\). This result implies that the concurrence of the Choi state sets an upper bound on the concurrence of the final state. Therefore, the concurrence of the Choi state defines the ability of its dual quantum channel to maintain entanglement.

This formalism can be applied directly to the spin–orbit QFC channel. From Eq.~\eqref{EQ_Channel_Raw}, we find
\begin{equation}
    \rho_{\mathcal{E}} =\frac{(\mathcal{I}\otimes  K^\dagger)\lvert\Phi^+)(\Phi^+\rvert(\mathcal{I}\otimes K)}{\textrm{Tr}[(\mathcal{I}\otimes  K^\dagger)\lvert\Phi^+)(\Phi^+\rvert(\mathcal{I}\otimes K)]} = \lvert\psi_\mathcal{E})(\psi_\mathcal{E}\rvert \,,
    \label{eq:ChoiMatrix}
\end{equation}
where $\lvert\Phi^+) = (1 \,  0\,  0\,  1)^\textrm{T}/\sqrt{2}$ is the computational basis representation of $|\Phi_+\rangle$, and where
\begin{equation}
    \lvert\psi_\mathcal{E}) = \frac{\sum_{ij=1}^2(K^\dagger)_{ji} \lvert i )\lvert j)}{\sqrt{\textrm{Tr}(KK^\dagger)}} \,.
\end{equation}
From Eq.~\eqref{eq:ChoiMatrix}, we see that the Choi state of the QFC channel is pure, and hence its concurrence can be calculated simply  as~\cite{Hill1997}
\begin{equation}
  C\bigl(\hat{\rho}_{\mathcal{E}}\bigr)
  =
  \frac{2|\det K^\dagger|}{\textrm{Tr}(K^\dagger K)} .
\end{equation}
At this point, it is convenient to write the matrix \(A\) in its singular-value decomposition
\(
   A = U\,\Sigma\,V^{\dagger},
\)
where \(U\) and \(V\) are unitary matrices and
\(\Sigma=\operatorname{diag}\!\bigl(\sigma_{-},\sigma_{+}\bigr)\)
collects the non-negative singular values \(\sigma_{\pm}\ge 0\).
It is straightforward to show that these singular values can be expressed as
\begin{equation}
    \sigma_{\pm}
    =
    \frac{\sqrt{1\pm\sqrt{\,1-C_{D}^{\,2}}}}{2},
\end{equation}
where \(C_{D}\) is the concurrence of the drive field as defined in Eq.~\eqref{eq:CoherenceDrive}. Consequently, the matrix that describes the map can be written in a convenient form
 \[
  K^\dagger =
  U\,\Sigma\,V^{\dagger}\,
  V\,\Sigma^{-1}V^{\dagger}\,
  V\,\sin\!\bigl(\kappa t\Sigma \bigr)\,V^{\dagger}
  =
  U\,\sin\!\bigl(\kappa t\Sigma \bigr)\,V^{\dagger}\,,
\]
from which we calculate the following expression for the concurrence of the Choi matrix   
\begin{equation}
    C(\hat{\rho}_{\mathcal{E}}) = \frac{2\left|\sin\!\left(\sigma_+\kappa t\right)
 \sin\!\left(\sigma_-\kappa t\right)\right|}
 {\sin^2\!\left(\sigma_+\kappa t\right)
 +\sin^2\!\left(\sigma_-\kappa t\right)}.
\label{ConcurenceChoiMatrix}
\end{equation}
Therefore, the concurrence depends non-trivially on the drive’s concurrence and the conversion efficiency.

In the regime of low conversion efficiency \(\kappa t \ll 1\), an intuitive interpretation of the QFC dynamics can be obtained. As $K^\dagger \xrightarrow[]{\kappa t \ll 1} \kappa t A $, Eqs.~\eqref{eq:CoherenceDrive} and~\eqref{eq:ChoiMatrix} lead to the following result
\begin{equation}
    \rho_\mathcal{E}\xrightarrow[]{\kappa t \ll 1}\rho_D\,.
\end{equation} 
In this case,
the result of Konrad \emph{et al.} presented in Eq.~\eqref{Konrad} implies that
\begin{equation}
  C\!\bigl(\hat{\rho}\bigr)
  \;\le\;
  C\!\bigl(\hat{\rho}_{\mathcal{E}}\bigr)\,
  C\!\bigl(\hat{\rho}_{0}\bigr)
  \;\xrightarrow[]{\kappa t \ll 1}\;
  C_D\,
  C\!\bigl(\hat{\rho}_{0}\bigr),
  \label{eq:ConcurrenceInequality}
\end{equation}
that is, the concurrence associated with the classical non-separability imposes a limiting factor on the entanglement evolution for the hybrid final state.  This provides an interesting instance in which the intrinsic correlations of the classical state are necessary—but not sufficient—for transferring genuinely quantum correlations.

\section{Detailed experimental setup}
\label{Detailed experimental setup}

\begin{figure}
    \centering
    \includegraphics[width=1.0\linewidth]{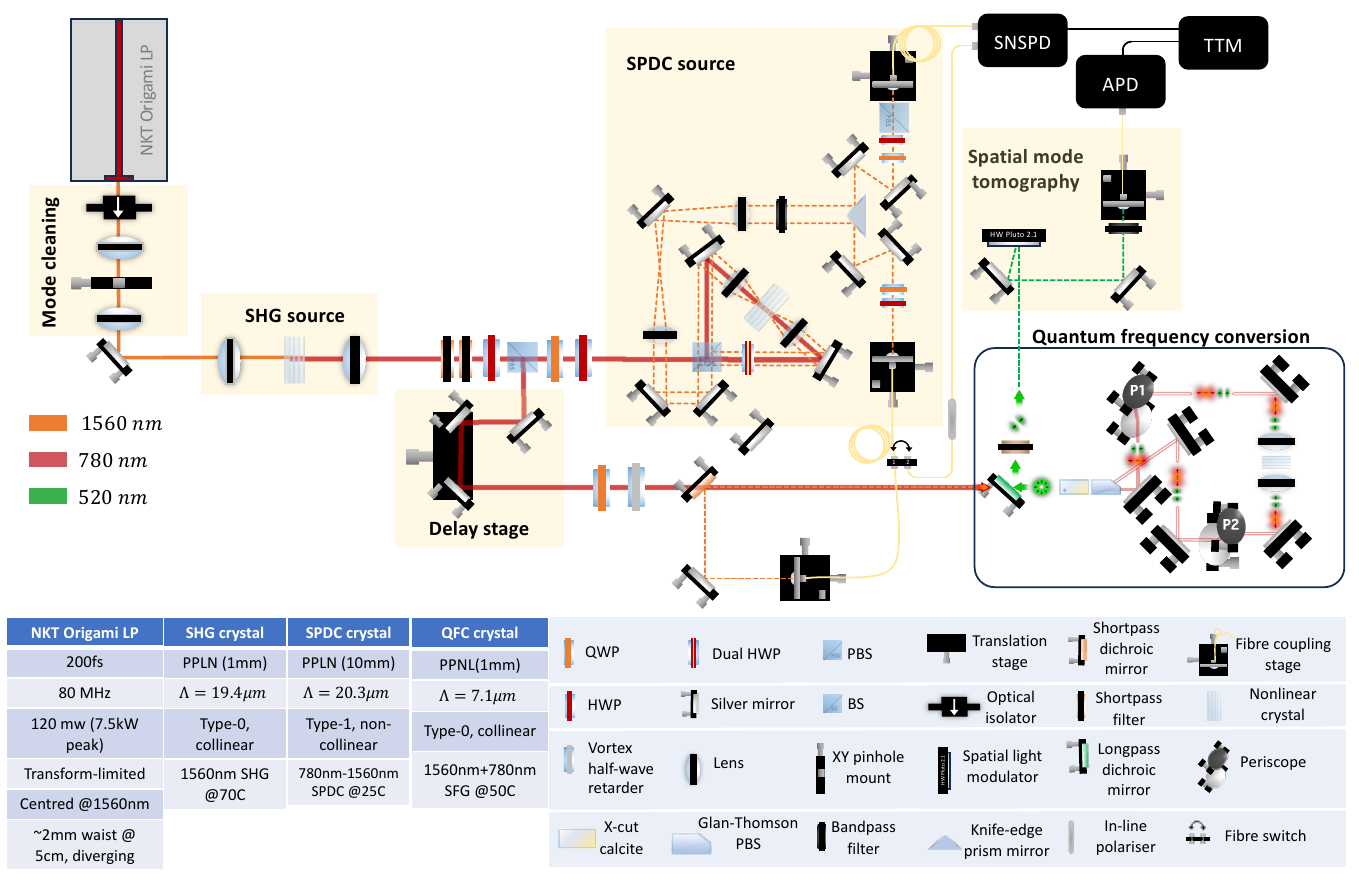}
    \caption{\textbf{Detailed experimental setup.} The figure shows each part of the experimental setup in full detail. Different wavelengths are represented in different colours, namely orange for 1560~nm, red for 780~nm, and green for 520~nm. The dashed lines coming from the photon pair source and the quantum frequency conversion part represent light fields on the single-photon level.}
    \label{FigSup_Setup}
\end{figure}

In Fig.~\ref{FigSup_Setup}, we show the expanded version of the experimental setup presented in the main text. Our laser source is a mode-locked laser from NKT Photonics, model Origami LP, centred at 1560~nm wavelength, with 120~mW average power, 220~fs pulse duration, and 80~MHz repetition rate. The laser is coupled to free space, diverging, and with some astigmatism. To obtain a better spatial profile, we clean the mode by focusing it through a $50~\mu$m pinhole with a plano-convex lens ($f=50~$mm). To generate the 780~nm laser required for both pumping the photon pair source and driving the quantum frequency conversion, we use the process of second-harmonic generation (SHG). For that, we focus the beam with an aspheric lens ($f=13.4~$mm) through a 1~mm-long periodically-poled lithium niobate crystal (PPLN, HC Photonics) with poling period $\Lambda_{\textrm{SHG}} = 19.4~\mu$m, which is phase-matched for SHG at a temperature of $70^\circ$C. The crystal length was limited to 1~mm to avoid temporal walk-off, keeping the SHG pulse duration to a minimum. The maximum average power obtained for the SHG was $37~$mW, which was maintained throughout the execution of the experiment.

After filtering out the fundamental beam with short-pass filters, we divide the SHG into two parts using waveplates and a polarising beamsplitter (PBS). The division is unbalanced - 2~mW are used to pump the SPDC source, and the remaining 35~mW are used to drive the QFC. To prepare the drive field, we use a combination of a quarter-wave plate and a half-wave vortex retarder (Thorlabs WPV10L-1550), as detailed in the main text. The vortex retarder prepares a spin-orbit laser field with a concurrence that is given by the degree of circular polarisation of the input. Therefore, we tune the concurrence of the drive field by rotating the quarter-wave plate before the vortex retarder. The prepared drive field passes through a short-pass dichroic mirror and is directed to the quantum frequency conversion part of the setup. 

\subsection{Source of polarisation-entangled photons}
\label{Source of polarisation-entangled photons}

The photon pair source employs a second PPLN crystal for spontaneous parametric downconversion (SPDC). The crystal has a poling period of $\Lambda_{\textrm{SPDC}} = 20.3~\mu$m and is operated at room temperature, yielding phase-matching when the pump polarisation is extraordinary ($e$) and the photon pairs are produced with ordinary ($o$) polarisation. To obtain a polarisation-entangled pair of photons from the type-I interaction, we use the Sagnac configuration shown in Fig~\ref{FigSup_Setup}. The design consists of assembling the crystal inside a Sagnac loop built with a PBS and a dual-wavelength half-wave plate (DHWP) rotated by $45^\circ$.  The SPDC process goes as follows. The $ e$-polarised pump field is reflected by the PBS, pumping the crystal directly and producing $ o$-polarised photon pairs. As the photons pass through the DHWP, their polarisations are rotated to $e$ and they exit the loop by reflecting from the PBS. On the other hand, the $ o$-polarised pump field is transmitted by the PBS, turning to $ e$-polarised by the DHWP before pumping the crystal. The photon pairs produced are $ o$-polarised and suffer no polarisation rotation, exiting the loop through the transmitted port of the PBS. Thus, if the pump field is in a balanced superposition of the $e$- and $ o$-polarisations, and the two exit ports are properly aligned, a polarisation-entangled state is obtained. 

In our case, we used the SPDC source in a slightly non-collinear configuration. By changing the crystal temperature, we change the phase-matching condition so that the photons are produced at an opening angle of approximately $2^\circ$. This change allows us to split the photons at the Fourier plane of the crystal with a knife-edge prism, as we show in Fig.~\ref{FigSup_Setup}. In the figure, the photon that goes upwards is the heralding photon, which is projected onto a chosen polarisation and measured directly with superconducting nanowire single-photon detectors (SNSPD, single-quantum). The second photon passes through motorised wave plates and is coupled to a single-mode fibre (SMF). This photon has two possible paths, accessible through a switchable fibre mating sleeve. When characterising the source, we project the photon's polarisation with an inline polariser and detect it with the SNSPDs, using the motorised waveplates to choose the polarisation projections. For the upconversion, we switch the fibre connector to send the photon directly to the upconversion stage, using the motorised waveplates to undo the polarisation transformation caused by the fibre. This arrangement ensures that the input state for upconversion matches the measured state of the SPDC source, without any effect of misalignment by sliding in and out a PBS before the fibre coupling. Lastly, the outcoupled photon is aligned to the drive field using a short-pass dichroic mirror, and both are directed to the upconversion stage.

\subsection{Sagnac-based quantum frequency conversion}
\label{Sagnac-based quantum frequency conversion}

For the quantum frequency conversion, we use the second Sagnac loop shown in Fig.~\ref{FigSup_Setup}. In this case, however, three different wavelengths must be compatible with the polarisation optics within the loop: the 1560~nm input photon, the 780~nm drive field, and the upconverted 520~nm photon. To meet these requirements, we use the achromatic Sagnac loop first introduced in Ref.~\cite{hentschel2009three} for three-colour SPDC sources. For the PBS, we use a calcite Glan-Thomson PBS (FocTek) preceded by a calcite crystal of the same length and rotated by $90^\circ$ around the propagation axis, ensuring that the polarisations are split without temporal distinguishability. The output facets of the PBS are cut at an angle, so that the beam paths are normal to the surface and the different colours are not misaligned by refraction. For the achromatic polarisation rotation, we use a periscope that takes the beam path upwards and rotates it by $90^\circ$, rotating not only the polarisation, but also the spatial modes involved. For the upconversion, we tightly focus the fields with uncoated achromatic lenses ($f=18.4$~mm) onto a 1~mm-long type-0 PPLN crystal, with poling period $\Lambda_{\textrm{QFC}} = 7.1~\mu$m and operated at $50^\circ$C. 

Finally, the upconverted photon is outcoupled using a long-pass dichroic mirror and projected onto the diagonal polarisation, erasing the distinguishability between the two loop directions at the cost of reducing the efficiency by 50\%. The tomography of the upconverted spatial mode is performed with a spatial light modulator and coupling to an SMF, as we detail in the following subsection, followed by the detection with an avalanche photodiode single-photon detector (APD, Laser Components Count-T).

\subsection{Spatial-mode tomography}
\label{Spatial-mode tomography}

We perform spatial-mode projections using the phase-flattening technique~\cite{bouchard2018measuring}, which consists of adding a phase pattern to the input field using a spatial light modulator (SLM) and coupling it to an SMF. For the first-order spatial modes, the coupling to the SMF is maximal when their phase patterns are correctly erased or flattened by the SLM, and is minimal when the SLM adds the phase of an orthogonal mode. Therefore, the SMF coupling efficiency measures the projection of the input field onto the mode whose conjugate phase pattern is displayed by the SLM. Typically, an extinction ratio of 20~-~30~dB is achieved between orthogonal mode projections, although this figure is extremely sensitive to alignment.

\section{Requirements for pulsed upconversion}
\label{Requirements for pulsed upconversion}

In our experiment, we use ultrashort pulses to drive the upconversion, aiming to leverage a high peak power to enhance the conversion efficiency. However, for the QFC to occur efficiently in the pulsed regime, the drive field and the input photons should arrive simultaneously at the nonlinear crystal and with good temporal overlap. In this context, it must be observed that, generally, SPDC photons do not have a well-defined temporal shape. Due to stringent phase-matching conditions and the energy conservation rule, photon pairs generally feature spectral-temporal entanglement. Consequently, a coherent spectrum can only be observed when the photons are measured in correlation, while, if measured alone, a single photon will be in a partially coherent superposition of its spectral components. As an ultrashort pulse requires a coherent spectrum, pulsed QFC is optimal for frequency-separable photons.

\begin{figure}
    \centering
    \includegraphics[width=0.8\linewidth]{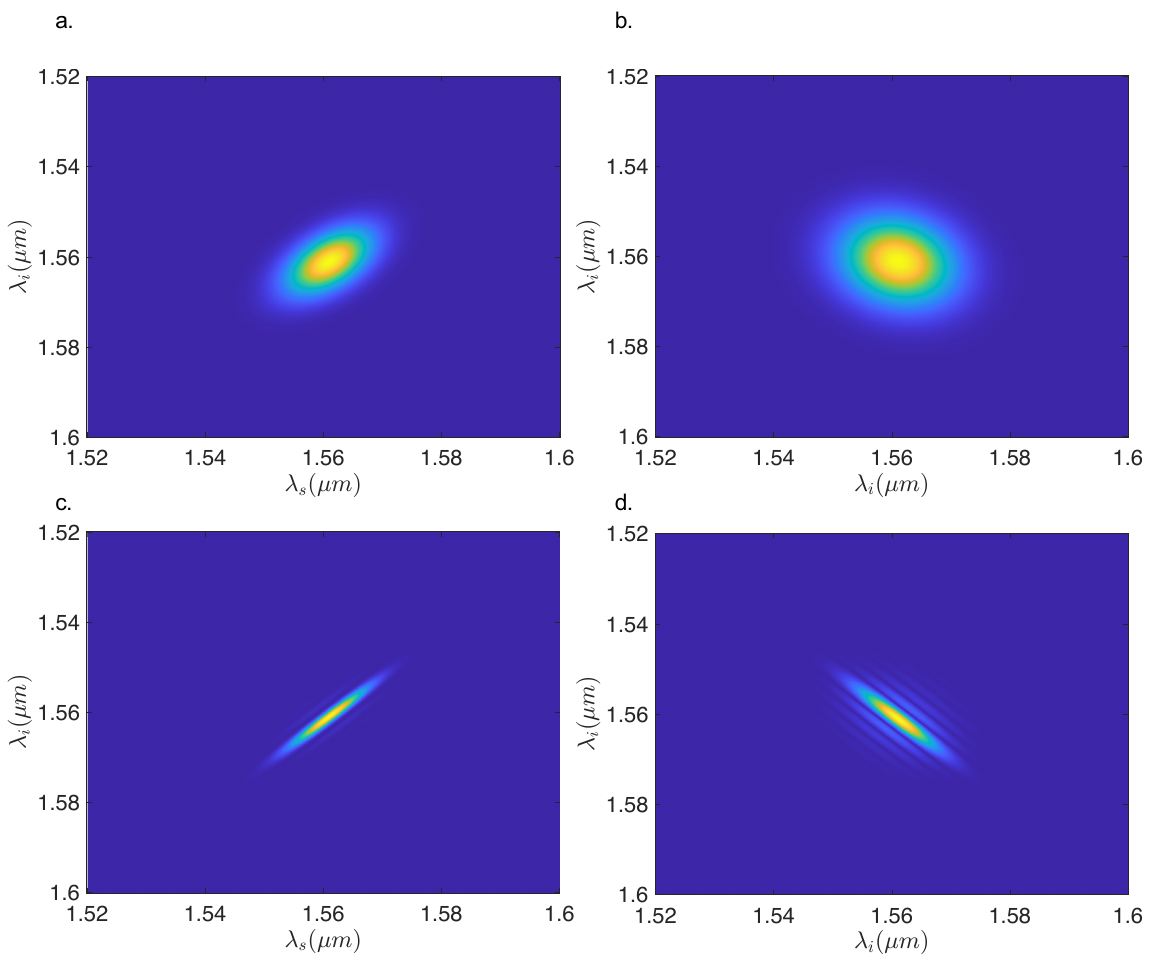}
    \caption{\textbf{Joint spectral analysis of the SPDC state.} (a) shows the absolute value of the joint spectral amplitude (JSA) for the SPDC state generated from a 10mm-long type-I PPLN crystal, with $\lambda_s$ and $\lambda_i$ denoting the wavelengths of the signal and idler photons, respectively. The pump field has a central wavelength of 780~nm and a Gaussian pulse profile with 220~fs pulse duration, and a Gaussian bandpass filter with 12~nm FWHM is applied to both the signal and idler wavelengths. (b) presents the absolute value of the reduced density matrix for the idler photon, which exhibits a purity of $85.9\%$. For comparison, (c) and (d) show the JSA and the idler state for a 10~mm-long type-0 PPLN crystal, also considering a 12~nm bandpass filter. In this case, the idler state has a purity of $18.4\%$.}
    \label{FigSup_JSA}
\end{figure}

In our experiment, we break the spectral entanglement of the SPDC state by using the type-I PPLN interaction, leveraging the group velocity matching between the 780~nm pump and the 1560~nm SPDC photons~\cite{yu2002broadband,horoshko2024few}. The group velocity matching enhances the phase matching bandwidth, so the whole pump spectrum contributes to the SPDC state. The result is a Joint Spectral Amplitude (JSA) that can be made separable with loose spectral filtering, as we show in Fig.~\ref{FigSup_JSA}.  In Fig.~\ref{FigSup_JSA}a, we show the calculated absolute value of the JSA for the photon pairs generated in our experiment, considering a 10~mm-long type-I PPLN crystal followed by a bandpass filter with 12~nm FWHM centred at 1560~nm. From the JSA, we calculate the reduced single-photon density matrix shown in Fig.~\ref{FigSup_JSA}b, which features a spectral purity is of $85.9\%$. As a rule of thumb, if the JSA has a symmetric Gaussian shape, the photon pairs are separable, and the reduced density matrix also has a Gaussian shape, meaning that the two photons have independently pure/coherent spectra. For comparison, we show in Figs.~\ref{FigSup_JSA}c and ~\ref{FigSup_JSA}d the JSA and reduced density matrix calculated for a 10~mm-long type-0 PPLN, which leads to single-photon spectral purity of only $18.4\%$. 

\begin{figure}
    \centering
    \includegraphics[width=0.4\linewidth]{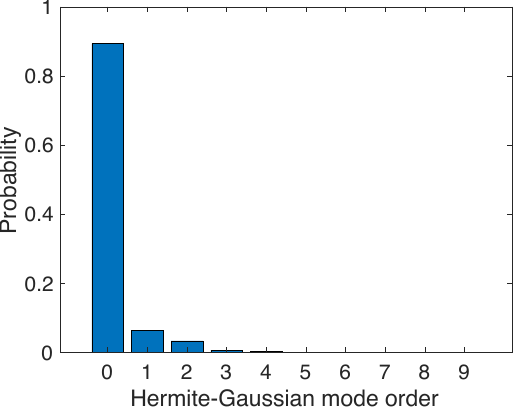}
    \caption{\textbf{Hermite-Gaussian decomposition of the unheralded single-photon state}. The figure shows the probability of detecting a single photon from the SPDC source used in our experiment (state is shown in Fig.\ref{FigSup_JSA}b) in each temporal Hermite-Gaussian mode. The mode duration is assumed to be 220~fs, matching the pulse duration of the pump field. }
    \label{FigSup_rhoImodes}
\end{figure}

To estimate the overlap between the drive field and the input photon during upconversion, we decompose the reduced density matrix in the temporal Hermite-Gaussian basis~\cite{raymer2020temporal}. We consider a mode basis with the same 220~fs duration as the drive field, obtaining the result shown in Fig.~\ref{FigSup_rhoImodes}. In our case, we find that the probability for the measured photon to be in the fundamental Gaussian spectral mode is $89.4\%$, which we take as a measure of the temporal overlap between the drive field and the waveform of the input photon. 

\begin{figure}
    \centering
    \includegraphics[width=0.7\linewidth]{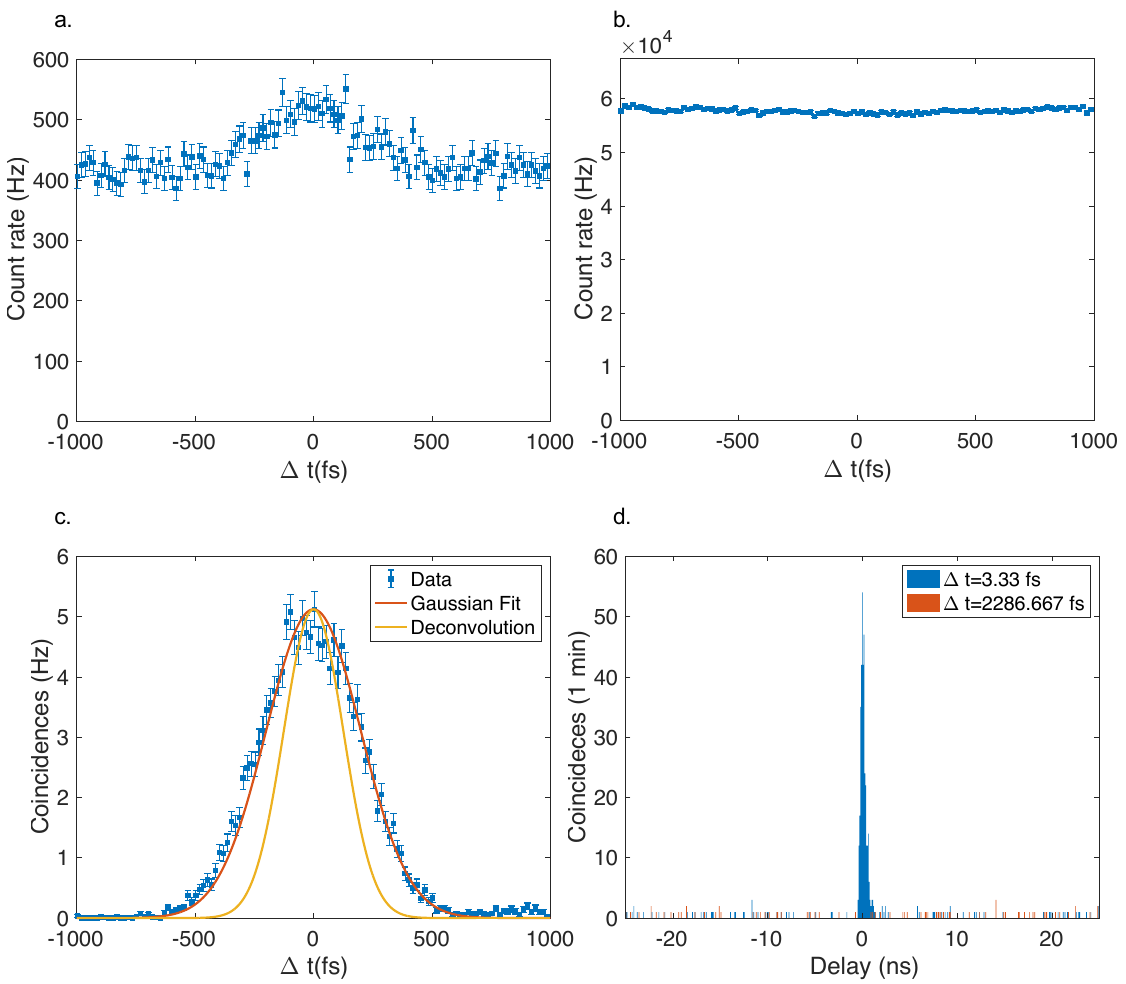}
    \caption{\textbf{Measurements of upconverted photons and temporal characteristics of the input photon.} (a) and (b) show the single-count rates for the upconverted and heralding photons, respectively, as a function of the time delay $\Delta t$ between the input photon and the drive pulse. (c) presents the coincidence count rate (markers) along with a Gaussian fit to the data (red curve), demonstrating that background noise present in the single counts is effectively suppressed in the coincidence measurements. The yellow curve shows the deconvolution of the 220~fs Gaussian temporal profile of the drive pulse from the fitted data, providing an estimate of the input photon's temporal waveform. (d) shows a histogram of raw coincidence counts versus the time difference between detection events at the SNSPD and the APD, illustrating the typical signal-to-noise ratio observed in the experiment. Periodic peaks in the accidental coincidences, spaced by 12.5~ns, correspond to the repetition rate of the laser source.}
    \label{FigSup_Delay}
\end{figure}

Experimentally, we verify that the input photon has an ultrashort waveform by measuring the upconversion signal while changing the arrival time of the input photon at the crystal. For this measurement, we set the drive to a radially polarised field and the spatial mode projection to a HG$_{10}$ mode, obtaining the results in Figs.~\ref{FigSup_Delay}a-d. In the Figs.~\ref{FigSup_Delay}a and \ref{FigSup_Delay}b, we show the raw detection rates for both the upconverted and the heralding photons as a function of the time delay inserted in the input photon's path. These results show a background detection rate of approximately 400~Hz for the upconverted photons, and a $\approx 500~fs$-wide signal peak centred at the zero-delay. We associate the background noise photons with unwanted nonlinear processes, as for example the phase-mismatched SPDC pumped by the drive field, followed by phase-matched QFC. This hypothesis is supported by the fact that we could not suppress the noise by adding spectral filters to the setup. However, we conclude that the noise level in our experiment does not significantly affect the coincidence measurements, as can be seen from Fig.~\ref{FigSup_Delay}c. The coincidence histogram shown in Fig.~\ref{FigSup_Delay}d also confirms that an excellent signal-to-noise ratio is obtained near the zero delay, as expected since the inefficiency in the upconversion decreases true coincidence counts and accidental coincidences at the same rate.

From the measured single-count rates $r_{\textrm{H}} = 60~kHz$ and $r_{\textrm{Up}} = 100~Hz$ for the heralding and upconverted photons, respectively, we can estimate the efficiency $\eta$ of our upconversion stage. Considering efficiencies of $\eta_{\textrm{SNSPD}}=80\%$ for the SNSPDs, $\eta_{\textrm{APD}}=60\%$ for the APDs, and taking into account the $50\%$ loss for the polarisation projection of the upconverted photon, we obtain
\begin{equation}
    \eta = 2\frac{r_{\textrm{Up}}\eta_{\textrm{SNSPD}}}{r_{\textrm{H}}\eta_{\textrm{APD}}}\approx0.4\%\,,
    \label{eq:efficiency}
\end{equation}
which already includes the losses from the SLM and SMF coupling, as well as the inherent losses in all dichroic mirrors, uncoated lenses, and the achromatic PBS. The same calculation can be performed by comparing the coincidence rates before and after QFC. Considering the measured pair rate of 2.6~kHz at the source and 5~Hz after QFC, we obtain $\eta\approx 0.5\%$, in agreement with the result of Eq.~\eqref{eq:efficiency}.

\section{Violation of Bell's inequality}
\label{Violation of Bell's inequality}

In this Appendix, we discuss the direct violation of Bell's inequality in our experiment. We use the Clauser–Horne–Shimony–Holt (CHSH) form of Bell's inequality, which can be written as 
\begin{equation}
    |\mathcal{B}(\hat{\rho})| = |E(a,b) - E(a,b^\prime) + E(a^\prime,b) + E(a^\prime,b^\prime)| \leq 2\,,
    \label{EQ_CHSHPoly}
\end{equation}
where $a$, $b$, $a^\prime$ and $b^\prime$ represent different measurement bases. The quantities  
\begin{equation}
    E(a,b) = \frac{\textrm{Tr}\left[{(1-\sigma_x)N^{(ab)}}\right]}{\textrm{Tr}\left[{(1+\sigma_x)N^{(ab)}}\right]}\,,
\end{equation}
measure the correlations in the chosen bases, where $\sigma_x$ is a Pauli matrix, and $N_{ij}^{(ab)}$ ($i,j = \{0,1\}$) denotes the coincidence rates measured for one photon projected onto the state $|i\rangle$ of the basis $a$, and the other photon projected onto the state $|j\rangle$ of the basis $b$. The violation of the Bell-CHSH inequality is incompatible with local hidden-variable theories, and it is taken as a hard signature of quantum entanglement. 

\begin{figure}
    \centering
    \includegraphics[width=0.6\linewidth]{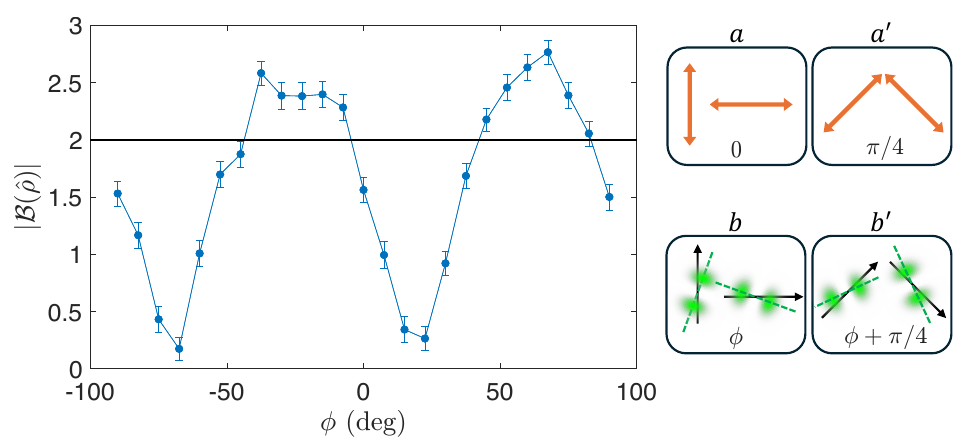}
    \caption{\textbf{Violation of the Bell-CHSH inequality with a hybrid two-colour entangled state.}  The figure shows measurements of the CHSH polynomial (Eq. \eqref{EQ_CHSHPoly}) for a quantum state shared between the polarisation of the heralding photon and the spatial mode of the upconverted photon. The measurements were performed using a radially polarised drive field, with the quantum state represented by the density matrix in Fig. 3b of the main text. Polarisation projections were performed in the computational and Hadamard bases, while the spatial mode projections were rotated by an angle $\phi$. Our results show a maximum of $|\mathcal{B}(\hat{\rho})| = 2.76\pm0.11$ at $67.5^\circ$, with the classical limit $|\mathcal{B}(\hat{\rho})|=2$ indicated by a solid line.}
    \label{FigSup_Bell}
\end{figure}

In our case, we choose the polarisation bases to be $a = \{|0\rangle,|1\rangle\}$, $a^\prime = \{\hat{R}(\pi/4)|0\rangle, \hat{R}(\pi/4)|1\rangle\}$,  $b = \{\hat{R}(\phi)|0\rangle, \hat{R}(\phi)|1\rangle\}$, and $b^\prime = \{\hat{R}(\phi+\pi/4)|0\rangle, \hat{R}(\phi+\pi/4)|1\rangle\}$, where $\hat{R}(\phi) = \exp{\left[i\phi\hat{Y}\right]}$ is an operator that rotates the state by an angle $\phi$ around the $y$ axis of the Bloch sphere. Physically, in our experiment, $a$ and $a^\prime$ represent polarisation projections along the horizontal/vertical and diagonal/anti-diagonal bases, while $b$ and $b^\prime$ represent projections onto HG$_{10}$ and HG$_{01}$ modes rotated clockwise by an angle $\phi$. We present our results in Fig. \ref{FigSup_Bell}, which shows a maximum value of $|\mathcal{B}(\hat{\rho})|_{\textrm{max}}=2.76\pm0.11$ for the CHSH polynomial, well above the classical limit $|\mathcal{B}(\hat{\rho})|=2$ and matching the Tsirelson bound of $2\sqrt{2}$~\cite{nielsen2010} within the standard deviation.

\clearpage

\bibliography{references}

\end{document}